\documentclass[prb,twocolumn,showpacs,superscriptaddress,superbib]{revtex4}

\usepackage{graphicx}
\usepackage{amsmath}
\usepackage{amsfonts}
\usepackage{bm}

\begin{document}

%......Title and other stuff
\title{Optical properties of self-organized wurtzite InN/GaN quantum dots: \\
A combined atomistic tight-binding and full configuration
interaction calculation}

\author{N. Baer}
\affiliation{Institute for Theoretical Physics, Semiconductor
Physics Group,
             University of Bremen,
             28334 Bremen, Germany}

\author{S. Schulz}
\author{S. Schumacher}
\affiliation{Institute for Theoretical Physics, Solid State Theory
Group,
             University of Bremen,
             28334 Bremen, Germany}

\author{P. Gartner}
\affiliation{Institute for Theoretical Physics, Semiconductor
Physics Group,
             University of Bremen,
             28334 Bremen, Germany}
\affiliation{National Institute of Materials Physics,
             PO Box MG-7,
             Bucharest-Magurele, Romania}

\author{G. Czycholl}
\affiliation{Institute for Theoretical Physics, Solid State Theory
Group,
             University of Bremen,
             28334 Bremen, Germany}
\author{F. Jahnke}
\affiliation{Institute for Theoretical Physics, Semiconductor
Physics Group,
             University of Bremen,
             28334 Bremen, Germany}

\date{\today}


\begin{abstract}

In this work we investigate the electronic and optical properties of
self-assembled InN/GaN quantum dots. The one-particle states of the
low-dimensional heterostructures are provided by a tight-binding
model that fully includes the wurtzite crystal structure on an
atomistic level. Optical dipole and Coulomb matrix elements are
calculated from these one-particle wave functions and serve as an
input for full configuration interaction calculations. We present
multi-exciton emission spectra and discuss in detail how Coulomb
correlations and oscillator strengths are changed by the
piezoelectric fields present in the structure. Vanishing exciton and
biexciton ground state emission for small lens-shaped dots is
predicted.

\end{abstract}

\maketitle

In recent years, semiconductor quantum dots (QDs) have been the
subject of intense experimental and theoretical research. As a new
material system, group-III nitride based devices are of particular
interest due to their wide range of emission frequencies from red to
ultraviolet and their potential for high-power electronic
applications \cite{Jain2000,Vurgaftman2003}. Being a technologically
promising system, we study self-assembled InN/GaN QDs, which are
typically grown by molecular beam epitaxy in Stranski-Krastanov
growth mode. A theoretical description of the one-particle states in
terms of a tight-binding (TB) model is presented, which provides a
powerful approach to the electronic states of low-dimensional
heterostructures on an atomistic level
\cite{Santropete2003,Schulz2005}. For the calculation of optical
absorption and emission spectra, full configuration-interaction
(FCI) calculations \cite{Barenco1995,Baer2004} are used to obtain a
consistent description of correlated many-particle states. The
calculation of dipole and Coulomb matrix elements from the TB
one-particle wave functions facilitates the combination of these two
approaches and allows us to investigate optical transitions between
the interacting many-particle states of a QD with parameters
obtained from a microscopic model. For the investigated small
lens-shaped InN/GaN QDs, we report a negligible exciton and
biexciton ground state emission whereas at higher excitation
conditions strong emission from three to six exciton complexes is
obtained.

We consider lens-shaped InN QDs, grown in (0001)-direction on top of
an InN wetting layer (WL) and embedded in a GaN matrix. Their
circular symmetry around the $z$-axis (diameter
$d=4.5\,\mathrm{nm}$, height $h=1.6\,\mathrm{nm}$) preserves the
intrinsic $C_{3v}$ symmetry of the wurtzite crystal. For the WL we
assume a thickness of one lattice constant. We apply a TB-model with
an $sp^3$ basis $|\alpha,\mathbf{R}\rangle$, i.e., one s-state
($\alpha=s$) and three p-states ($\alpha=p_x,p_y,p_z$) per spin
direction at each atom site $\mathbf{R}$. In contrast to most other
III-V and II-VI semiconductors, one can neglect spin-orbit coupling
and crystal-field splitting in InN and GaN
\cite{Wei1996,Vurgaftman2003}. We include non-diagonal elements of
the TB-Hamiltonian matrix up to nearest neighbors and use the
two-center approximation of Slater and Koster \cite{Slater1954}
which yields 9 independent TB-parameters. These parameters are
empirically determined such that the characteristic properties of
the bulk bandstructure \cite{Fritsch2004,Zhao1999} in the vicinity
of the $\Gamma$ point are reproduced. With these TB-parameters, the
QD is modeled on an atomistic level where the parameters for each
site are set according to the occupying atoms (N, In, Ga). At the
InN/GaN interfaces averages of the parameters are used and effects
of the surfaces in the finite-size supercell are removed
\cite{Sapra2004}. The spontaneous polarization in the wurtzite
crystal structure lies within growth direction:
$\mathbf{P}=P\mathbf{e}_z$. Additionally, a strain-induced
piezoelectric field occurs that is quite strong for the investigated
InN/GaN heterostructures. The piezoelectric field is determined by
solution of the Poisson equation. The strain contribution to the
polarization is approximated in a way following
Ref.~\onlinecite{Derinaldis2002}. For our chosen dot geometry, even
a more sophisticated inclusion of strain effects \cite{Andreev2001}
will generate merely small lateral contributions to the
piezoelectric field \cite{Saito2002}, which are therefore neglected
in the following. Lattice mismatch parameters and strain tensors are
taken from Ref.~\onlinecite{Shi2003}, the small thermal strain
contribution is neglected \cite{Derinaldis2002}. The calculation
yields a reasonable value of $5.5\,\mathrm{MV/cm}$ for the electric
field inside the QD. The resulting electrostatic potential is
included in the TB model as a site-diagonal potential
$V_p(\mathbf{r})=-e_0\phi_p(\mathbf{r})$. This method has
successfully been applied to quantum well \cite{Sala1999} and QD
\cite{Saito2002} structures before. By including the piezoelectric
field the quantum confined Stark effect (QCSE) and its influence on
the Coulomb matrix elements and the oscillator strengths can be
studied.

\begin{figure}
\includegraphics[scale=0.68]{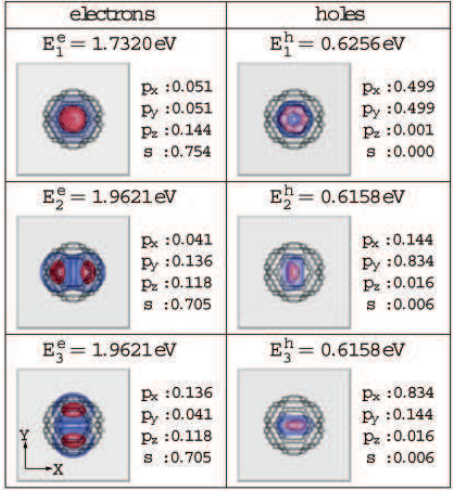}
\caption{\label{oneparticle}(Color online). The QD geometry is shown
from atop.  The structure is visualized and isosurfaces of the
charge density for the three energetically lowest electron (left)
and hole (right) states are included for $10\,\%$ (blue) and
$50\,\%$ (red) of the maximum value. For the holes the atomistic
structure of the wurtzite crystal becomes most apparent for the
excited p-states. The corresponding energies
($E_{1,2,3}^{\text{e,h}}$) of electron and hole states measured from
the valence band maximum of bulk GaN and the atomic orbital
character for each wave function are given. The dominant
contributions are highlighted.}
\end{figure}

The discussed QD confines three bound states for the electrons. They
are included in the FCI calculations, together with the three lowest
one-particle hole states that are spectrally well separated from the
other localized hole states. These three one-particle states for the
electrons and holes  and their energies are depicted in
Fig.~\ref{oneparticle}. As the lowest state for electrons and holes
is invariant under rotation by $2\pi/3$, it is denoted as s-state.
The two excited states are classified as p-states according to their
symmetry properties and their two-fold degeneracy. The hole states
show a strong band mixing visible in the orbital character
(Fig.~\ref{oneparticle}), which is in agreement with other
multi-band approaches \cite{Wei1996,Fonoberov2003}.

As emphasized above, a TB model represents an atomistic approach to
describe the electronic structure of low-dimensional
heterostructures. However, explicit knowledge about a basis set of
localized Wannier states is not required for the calculation of
one-particle energies and wave functions. Only the basic assumptions
about the localized (atomic) orbitals, i.e. symmetry, spatial
orientation \cite{Slater1954}, and orthogonality, enter the TB
Hamiltonian. Nevertheless, for the calculation of dipole and Coulomb
matrix elements one needs -- in principle -- the localized basis
states. For the Coulomb matrix elements, however, which are
dominated by long-range contributions, the explicit knowledge of the
atomic orbitals is in practice not required. This is because the
structure of the localized orbitals is of significance only for
on-site and nearest-neighbor interactions, which contribute less
than $5\,\%$ to the total Coulomb matrix elements. These findings
are in agreement with Ref.~\onlinecite{Sun2000}. Thus, the matrix
elements are approximated by a sum over the TB coefficients at atom
sites $\mathbf{R},\mathbf{R}'$ with orbital indices $\alpha,\beta$:
\begin{align}
V_{ijkl}=\sum_{\mathbf{R}\mathbf{R}'}\sum_{\alpha\beta}c_{\mathbf{R}\alpha}^{i\ast}
c_{\mathbf{R}'\beta}^{j\ast}c^k_{\mathbf{R}'\beta}c^l_{\mathbf{R}\alpha}\frac{e_0^2}{4\pi\varepsilon_0\varepsilon_r|\mathbf{R}-\mathbf{R}'|}
\,.
\end{align}
The labels $i,j,k,l$ refer either to electron or to hole states in
case of the repulsive electron-electron and hole-hole interaction,
or $i,l$ label electron and $j,k$ hole states for the attractive
electron-hole interaction. The considerably smaller matrix elements
of the electron-hole exchange interaction are neglected. The
electronic charge and the vacuum dielectric constant are denoted by
$e_0$ and $\epsilon_0$, respectively. We use the InN dielectric
constant $\varepsilon_{r}=8.4$ according to
Ref.~\onlinecite{Shi2003} since the wave functions are almost
completely confined inside the QD.

For the calculation of dipole matrix elements
$\mathbf{d}^{eh}_{ij}=e_0\langle\psi^e_i|\mathbf{r}|\psi^h_j
\rangle$, the explicit structure of the localized orbitals is
required as the dipole-operator has mainly local character. Standard
Slater orbitals \cite{Slater1930} have been used in earlier
calculations \cite{Lee2002} within orthogonal TB models. While they
include the correct symmetry properties, the missing orthogonality
limits their applicability. To overcome this problem, we use
numerically orthogonalized Slater orbitals. To properly treat the
slight non-locality of the dipole operator \cite{Whaley1998} and the
anion-cation structure of the crystal, the matrix elements are
calculated including up to second nearest neighbors. The only
relevant dipole matrix elements are
$\mathbf{e}\mathbf{d}^{eh}_{sp_x}=\mathbf{e}\mathbf{d}^{eh}_{sp_y}$
and
$\mathbf{e}\mathbf{d}^{eh}_{p_xs}=\mathbf{e}\mathbf{d}^{eh}_{p_ys}$,
where $\mathbf{e}=1/\sqrt{2}(1,1,0)$ denotes the light polarization
vector. All other matrix elements vanish due to the overall symmetry
of the connected one-particle states \cite{Bagga2005,Nirmal1995}.
The resulting optical selection rules are in strong contrast to what
is known from many other III-V and II-VI heterostructures and cannot
be explained within an one-component effective-mass approach
\cite{Li1999,Baer2004}.

\begin{figure}
\includegraphics[scale=0.88]{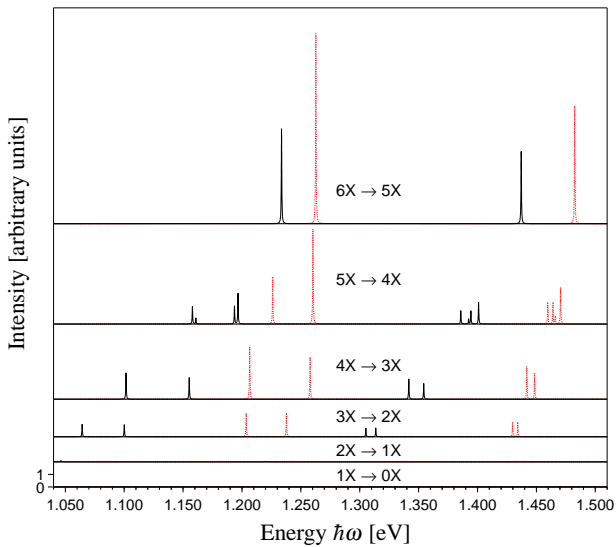}
\caption{\label{Fig2}(Color online). Emission spectra for a quantum
dot with different number of excitons, with (solid line) and without
(dotted line) piezoelectric field. Initial states are ground states
with total spin $z$-component $S_z=0$.}
\end{figure}

The single-particle states and Coulomb interaction matrix elements
are used in the FCI calculations to determine the multi-exciton
states. In a second step, Fermi's golden rule is evaluated for
dipole transitions between these Coulomb-correlated states in order
to obtain the multi-exciton emission spectra
\cite{Barenco1995,Baer2004}. The results for an initial filling of
the dot with one up to six excitons are depicted in Fig.~\ref{Fig2}
with (solid line) and without (dotted line) the piezoelectric field.
In the emission spectra two clusters of peaks are clearly visible,
one on the high energy side, $\hbar\omega>1.24\,\mathrm{eV}$, and
one on the low energy side, $\hbar\omega<1.3\,\mathrm{eV}$ (explicit
numbers given in the text refer to the results including the
piezoelectric field). As a characteristic feature, the high (low)
energy transitions originate from a recombination process involving
an electron (hole) in an excited state and a hole (electron) in the
ground state. As the dipole matrix element $d^{eh}_{ss}$ is
negligible and the exciton and biexciton ground states are dominated
by configurations where all the carriers are in the s-shell, the
corresponding transitions $1X\rightarrow0X$ and $2X\rightarrow1X$
remain dark. The gap between the two sets of clusters is, to a first
approximation, given by the difference in the involved one-particle
energies, which is then renormalized by the Coulomb interaction. In
a free particle picture each of the two clusters would collapse to
one line. The splitting within the clusters can be attributed to
transitions with different spin configurations of the final states.
These configurations are energetically separated by approximate
integer multiples of the Coulomb exchange integrals
$V^{eeee}_{sp_{x,y}sp_{x,y}}\approx 18\,\mathrm{meV}$ for the low
energy cluster, and $V^{hhhh}_{sp_{x,y}sp_{x,y}}\approx
4\,\mathrm{meV}$ for the high energy cluster. For clarification let
us consider the $3X\rightarrow2X$ transition. The $3X$ ground state
is dominated by configurations with both electrons and holes having
two carriers in the s- and one in the p-shell. Recombination of the
p-electron (hole) with the s-hole (electron) leaves behind two
electrons (holes) in the s-shell as well as one hole (electron) in
the s-shell and one in the p-shell. The latter can either form a
singlet or a triplet state, which leads to a splitting of around
$8\,\mathrm{meV}$ ($36\,\mathrm{meV}$), and is in accordance with
the FCI-calculation. As the number of e-h-pairs is increased one
observes an overall blue shift of the transitions. This shift is
less pronounced without the piezoelectric field and can be explained
in terms of the Hartree-Fock contributions to the interaction. The
in general somewhat larger peak height of the low-energy transition
is explained by the fact that the involved dipole matrix element
$d_{sp_{x,y}}^{eh}$ is larger than $d_{p_{x,y}s}^{eh}$, which enters
the high energy transitions.

The inclusion of the piezoelectric field for the strained wurtzite
crystal structure gives rise to a QCSE which creates a strong (about
$220\,\mathrm{meV}$) redshift of the one-particle gap energy.
Additionally, the Coulomb matrix elements are strongly modified and
the oscillator strengths are approximately halfed due to the spatial
separation of electron and hole wave functions.

In conclusion, we successfully combined two state-of-the-art
approaches, the atomistic tight-binding model and full
configuration-interaction calculations, to investigate the optical
properties of the technologically very promising InN/GaN QD system.
Multi-exciton emission spectra are calculated with microscopically
determined input parameters, which reveal the strong influence of
bandmixing effects on the optical transitions between the Coulomb
correlated many-particle states. As an important consequence for
future optoelectronic applications we predict vanishing exciton and
biexciton ground state emission for small lens-shaped InN/GaN QDs.

We acknowledge financial support by the Deutsche
Forschungsgemeinschaft and a grant for CPU time from the NIC at the
Forschungszentrum J\"ulich.

%\bibliography{../literature}
%\bibliographystyle{apsrev}

\end{document}